\documentclass[aps,prl,preprint,groupedaddress]{revtex4}
\usepackage{graphicx}

\begin{document}


\title{Corollary to the Hohenberg-Kohn Theorem}


\author{Xiao-Yin Pan and Viraht Sahni}
\affiliation{Department of Physics, Brooklyn College of the City
University of New York, 2900 Bedford Avenue, Brooklyn, New York
11210, and The Graduate School of the City University of New
York,360 Fifth Avenue, New York, New York 10016.}


\date{\today}

\begin{abstract}
 According to the Hohenberg-Kohn theorem, there is an invertible
  one-to-one relationship between the Hamiltonian $\hat{ H}$ of a system
  and the corresponding ground state density $\rho ({\bf r})$. The extension
   of the theorem to the time-dependent case by Runge and Gross states
   that there is an invertible one-to-one relationship between the
   density $\rho ({\bf r} t)$ and the Hamiltonian $\hat {H}(t)$. In the proof of the
   theorem, Hamiltonians $\hat{H}/\hat {H}(t)$ that differ by an \textit{\textbf{additive}}
   constant $C$/ function $C(t)$ are considered equivalent.
   Since the constant $C$/ function $C(t)$ is extrinsically additive, the
    physical system defined by these differing Hamiltonians $\hat{H}/\hat {H}(t)$
    is the \textit{\textbf{same}}. Thus, according to the theorem, the density $\rho ({\bf r} )/\rho ({\bf r} t)$
     uniquely determines the physical system as
     defined by its Hamiltonian $\hat{H}/\hat {H}(t)$.  Hohenberg-Kohn, and by
     extension Runge and Gross, did not however consider the case
      of a set of degenerate Hamiltonians $\{\hat{H}\}/\{\hat {H}(t)\}$ that differ
      by an \textit{\textbf{intrinsic}} constant $C$/function $C(t)$ but which represent
      \textit{\textbf{different}} physical systems and yet possess the same density
     $\rho ({\bf r} )/\rho ({\bf r} t)$ . The intrinsic constant C/function C(t) contains
      information about the different physical systems and  helps
      differentiate between them. In such a case,
       the density $\rho ({\bf r} )/\rho ({\bf r} t)$  cannot distinguish between these different
        Hamiltonians. In this paper we construct such a set of degenerate
         Hamiltonians $\{\hat{H}\}/\{\hat {H}(t)\}$. Thus, although the proof of the Hohenberg-Kohn
         theorem is \textbf{\emph{independent}} of whether the constant $C$/function $C(t)$ is additive
         or intrinsic, the applicability of the theorem is restricted to excluding
         the case of the latter. The corollary is as follows: Degenerate Hamiltonian
         $\{\hat{H}\}/\{\hat {H}(t)\}$ that represent different physical systems, but which
         differ by a constant  $C$/function $C(t)$ , and yet possess the same density
          $\rho ({\bf r} )/\rho ({\bf r} t)$, cannot be distinguished on the basis of the Hohenberg-Kohn/Runge-Gross theorem.

\end{abstract}

\pacs{}

\maketitle

\section{I. Introduction and Corollary}
  This paper provides further insight into the first of the two
  Hohenberg-Kohn (HK) theorems \cite{1} that constitute the
  rigorous mathematical basis for density functional theory.
  According to the  theorem, for a system of $N$ electrons in
  an external field $\mathcal{F}^{ext}({\bf r}) =-{\bf \nabla} v({\bf
  r})$, the \textbf{\textit{ground}} state electronic density $\rho ({\bf
  r})$ for a nondegenerate state determines the external potential
  energy $v({\bf r})$ uniquely to within an \textit{\textbf{unknown trivial
  additive constant }}$C$. Since the kinetic energy $\hat T$ and
  electronic-interaction potential energy $\hat U$ operators are
  known, the Hamiltonian $\hat H$ is explicitly known. The solutions $\Psi$ of the
  corresponding time-independent  Schr{\"o}dinger equation,  for both
  ground and excited states, then determine the properties of the
  electronic system. The wave function is thus a functional of the
  density: $\Psi =\Psi [\rho]$, and therefore all expectations are
  unique functionals of the density. Thus, the ground state
  density $\rho ({\bf r})$ determines \textit{\textbf{all}} the properties of the
  system. \\

     In the extension of the first HK theorem to the
  time-dependent case, Runge and Gross(RG) \cite{2} prove that for a
  system of $N$ electrons in a time-dependent external field $\mathcal{F}^{ext}({\bf r}t) =-{\bf \nabla} v({\bf
  r}t)$, such that the potential energy $v({\bf r}t)$ is Taylor-
  expandable about some initial time $t_{0}$, the density $\rho
  ({\bf r}t)$ evolving from some fixed initial state
  $\Psi(t_{0})$, determines the external potential energy uniquely
  to within an \textit{\textbf{additive purely time-dependent function}}
  $C(t)$. Again, as the kinetic and electron-interaction potential
  energy operators are already defined, the Hamiltonian ${\hat
  H}(t)$ is known. The solution $\Psi(t)$ of the time-dependent
  Schr{\"o}dinger equation then determines the system properties.
  Equivalently, the wave function $\Psi(t)$ is a functional of
  the density, unique to within a time-dependent phase. As such
  all expectation values are unique functionals of the density,
  the phase factor cancelling out. \\

     In the preamble to their proof, HK/RG consider Hamiltonians $\hat{H}/\hat {H}(t)$
that differ by an additive constant $C$/function $C(t)$ to be
equivalent. In other words, the \textit{\textbf{physical system}}
under consideration as defined by the electronic Hamiltonian
remains the \textit{\textbf{same}} on addition of this
constant/function which is \textit{\textbf{arbitrary}}. Thus,
measurement of properties of the system, other than for example
the total energy $E/E(t)$, remain invariant. The theorem then
proves that \textit{\textbf{each}} density $\rho({\bf
r})/\rho({\bf r} t)$ is associated with \textit{\textbf{one}} and
\textit{\textbf{only one}} Hamiltonian $\hat{H}/\hat {H}(t)$ or
physical system: the density $\rho({\bf r})/\rho({\bf r} t)$
determines that unique Hamiltonian $\hat{H}/\hat {H}(t)$ to within
an additive constant
$C$/function $C(t)$.\\

  HK/RG, however, did not consider the case of a set of Hamiltonians $\{\hat{H}\}/\{\hat
  {H}(t)\}$ that
  represent \textit{\textbf{different }} physical systems  which  differ by an \textit{\textbf{intrinsic}}
  constant $C$/function $C(t)$, but which yet have the \textbf{\textit{same}} density $\rho({\bf r})/\rho({\bf r}
  t)$. By intrinsic constant  $C$/function
  $C(t)$ we mean one that is inherent to the system and not
  extrinsically additive. Thus, this constant $C$/function $C(t)$
  helps distinguish between the different Hamiltonians in the set $\{\hat{H}\}/\{\hat
  {H}(t)\}$, and is consequently  \textit{\textbf{not arbitrary}}. That the physical systems are
  \textit{\textbf{different}} could, of course,  be confirmed by
  experiment. Further,  the density $\rho({\bf r})/\rho({\bf r}
  t)$
   would then not be able to distinguish between the different
 Hamiltonians $\{\hat{H}\}/\{\hat {H}(t)\}$ or physical systems, as it is the same for all of them.\\

  In this paper we construct a \textit{\textbf{set}} of model
  systems with \textit{\textbf{different }} Hamiltonians $\{\hat{H}\}/\{\hat {H}(t)\}$
  that differ by a constant $C$/function $C(t)$ but which
  \textit{\textbf{all}} possess the same density $\rho({\bf r})/\rho({\bf r}
  t)$. This is the Hooke's species: atom, molecule, all positive
  molecular ions with number of nuclei $\mathcal{N}$ greater than
  two. The constants $C$/function $C(t)$ contain information about
  the system, and are intrinsic to distinguishing between the
  different elements of the species.\\

  The corollary to the HK/RG theorem is as follows: Degenerate Hamiltonians $\{\hat{H}\}/\{\hat
  {H}(t)\}$ that differ by a constant $C$/function $C(t)$ but which
  represent different physical systems all possessing the same
  density $\rho({\bf r})/\rho({\bf r} t)$ cannot be distinguished on the basis of the HK/RG
  theorem. That is, for such systems,
   the density $\rho({\bf r})/\rho({\bf r} t)$ cannot determine each
  external potential energy $v({\bf r})/ v({\bf r} t)$, and hence
  each Hamiltonian of the set $\{\hat{H}\}/\{\hat{H}(t)\}$, uniquely.\\

       In the following sections, we describe the Hooke's species
    for the time-independent and time-dependent cases to prove the
    above corollary. \\
\pagebreak
\section{II. HOOKE'S SPECIES }

\textbf{A}.   ~\textbf{Time-Independent case}.\\

  Prior to describing the Hooke's species, let us consider the
  following Coulomb species of  two-electron systems and $\mathcal{N}$ nuclei: the
   Helium  atom ($\mathcal{N}=1$; atomic number $Z=2$), the Hydrogen molecule
   ($\mathcal{N}=2$; atomic number of each nuclei $Z=1$), and the
   positive molecular ions ($\mathcal{N}>2$; atomic number of each nuclei
$Z=1$).\\

  In atomic units, the Hamiltonian of the Coulomb species is
\begin{equation}
   \hat{H}_{\mathcal{N}} =\hat{T}+ \hat{U}+ \hat{V}_{\mathcal{N}}
   \end{equation}
   where $\hat {T}$ is the kinetic energy operator:
   \begin{equation}
   \hat{T}=-\frac{1}{2}\sum^{2}_{i=1}{\nabla_{i}}^2 ,
   \end{equation}
   $\hat {U}$ the electron-interaction potential energy operator:
   \begin{equation}
   \hat{U}=\frac{1}{|{\bf r}_{1}-{\bf r}_{2}|},
   \end{equation}
   and $\hat{V}_{\mathcal{N}}$ the external potential energy
   operator:
    \begin{equation}
    \hat{V}_{\mathcal{N}}=\sum ^{2}_{i=1} v_{\mathcal{N}}({\bf
    r}_{i})  ,
    \end{equation}
    with
     \begin{equation}
     v_{\mathcal{N}}({\bf r})=  \sum^{\mathcal{N}}_{j=1} f_{C}({\bf r}-{\bf R}_{j}).
     \end{equation}
     where
   \begin{equation}
  f_{C}({\bf r}-{\bf R}_{j})=-\frac{1}{|{\bf r}-{\bf R}_{j}|}.
   \end{equation}
     Here ${\bf r}_{1}$ and  ${\bf r}_{2}$ are positions of the
     electrons, ${\bf R}_{j}(j=1,...\mathcal{N})$  the positions of
     the nuclei, and $f_{C}({\bf r}-{\bf R}_{j})$ the Coulomb external potential energy function. Each element
     of the Coulomb species represents a \textit{\textbf{different}} physical system. ( The species could be further generalized by requiring   each nuclei to have a different charge.)\\

      Now suppose the ground state density $\rho({\bf r})$ of the
      Hydrogen molecule were known. Then, according to the HK
      theorem, this density uniquely determines the external
      potential energy operator to within an additive constant $C$:
 \begin{equation}
 {\hat V}_{\mathcal{N}=2}=-\frac{1}{|{\bf r}_{1}-{\bf R}_{1}|}-\frac{1}{|{\bf r}_{1}-{\bf
 R}_{2}|}-\frac{1}{|{\bf r}_{2}-{\bf R}_{1}|}- \frac{1} {|{\bf r}_{2}-{\bf
 R}_{2}|}.
 \end{equation}
 Thus, the Hamiltonian of the Hydrogen molecule is exactly known
 from the ground state density. Note that in addition to the functional form of
 the external potential energy, the density also explicitly defines
 the positions ${\bf R}_{1}$ and ${\bf R}_{2}$ of the nuclei.\\

  The fact that the ground state density determines the external
  potential energy operator, and hence the Hamiltonian may be
  understood as follows. Integration of the
  density leads to the number $N$ of the electrons: $\int \rho
  ({\bf r}) d {\bf r}=N$. The cusps in the electron density which satisfies the electron-nucleus cusp
  condition \cite{3}, determine in turn the positions of the $\mathcal{N}$
  nuclei and their charge $Z$. Thus, the external potential energy
  operator ${\hat V}_{\mathcal{N}}=\sum_{i} v_{\mathcal{N}}({\bf r}_{i})$, and therefore the
  Hamiltonian ${\hat H}$ are known.\\

  The Hooke's species comprise of two electrons coupled
  harmonically to a variable number $\mathcal{N}$ of nuclei. The
  electrons are coupled to each nuclei with a different spring
  constants $k_{j},j=1...\mathcal{N}$. The species comprise of the
  Hooke's atom ($\mathcal{N}=1$, atomic number $Z=2$, spring constant k), the Hooke's molecule
   ($\mathcal{N}=2$; atomic number of each nuclei $Z=1$, spring constants $k_{1}$ and $k_{2}$), and the Hooke's
   positive molecular ions ($\mathcal{N}>2$, atomic number of each nuclei $Z=1$, spring constants $k_{1}, k_{2},
   k_{3}...k_{\mathcal{N}}$). The Hamiltonian ${\hat
   H}_{\mathcal{N}}$ of this species is the same as that of the
   Coulomb species of Eq.(1) except that the external potential
   energy function is $f_{H}({\bf r}-{\bf R}_{j})$, where
\begin{equation}
f_{H}({\bf r}-{\bf R}_{j})=\frac{1}{2} k_{j}({\bf r}-{\bf
R}_{j})^{2}.
\end{equation}
Just as for the Coulomb species, each element of the Hooke's species
represents a \textbf{\textit{different}} physical system. Thus, for example, the
Hamiltonian for Hooke's atom is
\begin{equation}
    \hat{H}_{a}=-\frac{1}{2}{\nabla_{1}}^2-\frac{1}{2}{\nabla_{2}}^2+\frac{1}{|{\bf r}_{1}-{\bf
    r}_{2}|}+ \frac{1}{2} k[({\bf r}_{1}-{\bf R}_{1})^{2}+({\bf r}_{2}-{\bf
    R}_{1})^{2}],
    \end{equation}
and that of  Hooke's molecule is
\begin{equation}
    \hat{H}_{m}=-\frac{1}{2}{\nabla_{1}}^2-\frac{1}{2}{\nabla_{2}}^2+\frac{1}{|{\bf r}_{1}-{\bf
    r}_{2}|}+ \frac{1}{2}\{k_{1} [({\bf r}_{1}-{\bf R}_{1})^{2}+({\bf r}_{2}-{\bf
    R}_{1})^{2}]+k_{2} [
    ({\bf r}_{1}-{\bf R}_{2})^{2}+({\bf r}_{2}-{\bf R}_{2})^{2}]\},
    \end{equation}
where $k \neq k_{1} \neq k_{2} $,  and  so on for the various
Hooke's positive molecular ions   with $\mathcal{N}>2$.\\

 For the Hooke's species, however, the external potential energy operator
 ${\hat V}_{\mathcal{N}}$ which is
    \begin{equation}
    {\hat V}_{\mathcal{N}}=\frac{1}{2}\sum^{\mathcal{N}}_{j=1}[k_{j}({\bf r}_{1}-{\bf R}_{j})^{2}+k_{j}({\bf r}_{2}-{\bf R}_{j})^{2}],
     \end{equation}
   may be rewritten as
\begin{equation}
    {\hat V}_{\mathcal{N}}=(\frac{1}{2}\sum^{\mathcal{N}}_{j=1} k_{j})
    [({\bf r}_{1}-{\bf a})^{2}+({\bf r}_{2}- {\bf a})^{2}]+C(\{k\},\{ {\bf R} \},\mathcal{N}),
 \end{equation}
where the translation vector ${\bf a}$ is
\begin{equation}
{\bf a}=\sum^{\mathcal{N}}_{j=1} k_{j} {\bf R}_{j}/\sum^{\mathcal{N}}_{j=1} k_{j},
\end{equation}
   and the constant $C$ is
   \begin{equation}
   C=b-d
\end{equation}
with
  \begin{equation}
   b=\sum^{\mathcal{N}}_{j=1} k_{j} {\bf R}_{j}^{2},
\end{equation}
\begin{equation}
   d=(\sum^{\mathcal{N}}_{j=1} k_{j} {\bf R}_{j})^{2}/\sum^{\mathcal{N}}_{j=1} k_{j},
\end{equation}
or
\begin{equation}
   C=\frac{1}{2} \sum^{\mathcal{N}}_{i,j=1}{'} k_{i}k_{j}({\bf R}_{i}-{\bf R}_{j})^{2}/\sum^{\mathcal{N}}_{j=1} k_{j}.
\end{equation}
 From Eq.(12) it is evident that the Hamiltonians ${\hat H}_{\mathcal{N}}$ of the Hooke's species are those of a Hooke's atom ($\sum^{\mathcal{N}}_{j=1} k_{j}=k $), (to within a constant $C(\{k\},\{{\bf R}\},\mathcal{N})$), whose center of mass is at ${\bf a}$. The constant
 $C$ which depends upon the spring constants $\{k\}$, the positions of the nuclei $\{{\bf R}\}$, and the number $\mathcal{N}$ of the nuclei, differs from a trivial additive constant
 in that it is an \textit{\textbf{intrinsic}} part of \textit{\textbf{each}} Hamiltonian
 ${\hat H}_{\mathcal{N}}$, and distinguishes between the different elements of the species.
 It does so because the constant $C(\{k\},\{\bf R\},\mathcal{N})$ contains physical information about the system such as the positions $\{\bf R\}$ of the nuclei.\\

 Now according to the HK theorem, the ground state density determines the external potential
 energy, and hence the Hamiltonian, to within a constant. Since the density of \textit{\textbf{each}} element of the Hooke's species is that of the Hooke's atom, it  can only determine the Hamiltonian of a Hooke's atom and not the constant $C(\{k\},\{{\bf R}\},\mathcal{N})$. Therefore, it cannot determine the Hamiltonian  ${\hat H}_{\mathcal{N}}$
 for $\mathcal{N}>1$. This is reflected by the fact that the density of the elements of the Hooke's species does not satisfy the electron-nucleus cusp condition.( It is emphasized
 that although the `degenerate Hamiltonians' of the Hooke's species have a ground state
 wave function and density that corresponds to that of a Hooke's atom, each element of the species represents a \textit{\textbf{different}} physical system. Thus, for example, a neutron diffraction
 experiment on the Hooke's molecule and Hooke's positive molecular ion would \textit{\textbf{all}} give different results).\\

 It is also possible to construct a Hooke's species such that the density of each element
 is the \textit{\textbf{same}}. This is most readily seen for the case when the center
 of mass is moved to the origin of the coordinate system, i.e. for ${\bf a}=0$. This requires, from Eq.(13), the product of the spring constants and the coordinates of the nuclei
 satisfy the condition
 \begin{equation}
   \sum^{\mathcal{N}}_{j=1} k_{j} {\bf R}_{j}=0,
\end{equation}
 so that the external potential energy operator is then
 \begin{equation}
    {\hat V}_{\mathcal{N}}({\bf r})=\frac{1}{2}\sum^{\mathcal{N}}_{j=1}k_{j}{\bf r}^{2}+\frac{1}{2}\sum^{\mathcal{N}}_{j=1} k_{j}{\bf R}_{j}^{2},
     \end{equation}
 where ${\bf r}$ is the distance to the origin. If the sum $\sum^{\mathcal{N}}_{j=1} k_{j}$ is
 then adjusted to equal a particluar value of the spring constant $k$ of Hooke's atom:
 \begin{equation}
   \sum^{\mathcal{N}}_{j=1} k_{j}=k,
\end{equation}
 then the Hamiltonian ${\hat H}_{\mathcal{N}}$ of any element of the species may be rewritten
 as
\begin{equation}
   {\hat H}_{\mathcal{N}}(\{k\},\{{\bf R}\},\mathcal{N})= {\hat H}_{a}(k)+C(\{k\},\{{\bf R}\},\mathcal{N}) ,
\end{equation}
where ${\hat H}_{a}(k)$ is the Hooke's atom Hamiltonian and the constant $C(\{k\},\{\bf R\},\mathcal{N})$ is
\begin{equation}
  C(\{k\},\{{\bf R}\},\mathcal{N})=\sum^{\mathcal{N}}_{j=1} k_{j} {\bf R}_{j}^{2}.
\end{equation}
The solution of the Schr{\"o}dinger equation and the corresponding density for
\textit{\textbf{each}} element of the species are therefore the \textit{\textbf{same}}.\\

 As an example, again consider the case of Hooke's  atom  and molecule. For Hooke's atom $\mathcal{N}=1,{\bf R}_{1}=0$, and
     let us assume $k=\frac{1}{4}$. Thus, the external potential
     energy operator is
      \begin{equation}
      v_{a}({\bf r})=\frac{1}{2} k  r^{2}= \frac{1}{8} r^{2}.
      \end{equation}
      For this choice of $k$, the singlet ground state solution
      of the time-independent Schr{\"o}dinger equation $(\hat
      {H}_{\mathcal{N}} \Psi=E_{\mathcal{N}} \Psi)$ is analytical
      \cite{4}:
     \begin{equation}
     \Psi({\bf r}_{1} {\bf r}_{2})=D e^{-y^{2}/2} e^{-r^{2}/8}
     (1+r/2),
      \end{equation}
      where $ {\bf r}={\bf r}_{1}-{\bf r}_{2}, {\bf y}=({\bf r}_{1}+{\bf
      r}_{2})/2$, and $D= 1/[ 2 \pi^{5/4}(5 \sqrt{\pi}+8)^{1/2}]$.
      The corresponding ground state density $\rho({\bf
      r})=\langle \Psi |\hat {\rho} ({\bf r})|\Psi\rangle$, $\hat {\rho} ({\bf
      r})=\sum ^{2}_{i=1} \delta ({\bf r}-{\bf r}_{i})$ is
      \cite{5,6}
      \begin{equation}
     \rho({\bf r})=\frac{\pi \sqrt{2 \pi}}{r}D^{2} e^{-r^{2}/2} \{7
     r+ r^{3}+ (8/\sqrt{2\pi})r e^{-r^{2}/2} + 4 (1+r^{2})
     erf(r/\sqrt{2}\},
     \end{equation}
     where
       \begin{equation}
       erf(x)=\frac{2}{\sqrt{\pi}} \int ^{x}_{0} e^{-z^{2}} dz.
        \end{equation}
    For the Hooke's molecule, $\mathcal{N}=2, {\bf R}_{1}=- {\bf
    R}_{2}$, and we choose $k_{1}=k_{2}= \frac{1}{8}$, so that the external
    potential energy operator is
      \begin{equation}
      v_{m}({\bf r})=\frac{1}{8} r^{2}+
      \frac{1}{16}(R_{1}^{2}+R_{2}^{2})= \frac{1}{8} r^{2}+\frac{1}{8}
      R^{2},
      \end{equation}
    where $|{\bf R}_{1}|=R$. Thus, the Hamiltonian for Hooke's
    molecule differs from that of Hooke's atom by only the
    constant $\frac{1}{8} R^{2}$, thereby leading to the \textit{\textbf{same
    }}ground state wave function and density. However, the ground
    state energy of the two elements of the species differ by  $\frac{1}{8}
    R^{2}$.\\

    The above example demonstrating the equivalence of the
         density of the Hooke's atom and molecule is for a
         specific value of the spring constant $k$ for which the
         wave function happens to be analytical. However, this
         conclusion is valid for arbitrary value of $k$ for which
         solutions of the Schr{\"o}dinger equation exist but are
         not necessarily analytical.
        For example, if we assume that for each element of the
        species ($\mathcal{N}\geq 2$), all the spring constants
        $k_{j}, j=1,2,...\mathcal{N}$ are the same and designated
        by $k'$,  then for the
         three values of $k$ for the
         Hooke's atom corresponding to $k=\frac{1}{4}, \frac{1}{2},1$, the
         values of $k'$ for which the Hooke's molecule and
         molecular ion ($\mathcal{N}=3$) wave functions are the same are
         $k'=\frac{1}{8},  \frac{1}{12}; \; k'=\frac{1}{4},
         \frac{1}{6};\; k'=\frac{1}{2}, \frac{1}{3}$, respectively.\\

  Thus, for the case where the elements of the Hooke's species are all made to
  have the \textit{\textbf{same}} ground state density $\rho({\bf r})$, the density
  cannot, on the basis of the HK theorem, distinguish bewteen the different physical
  elements of the species. \\

  \textit{\textbf{Corollary}}: Degenerate time-independent Hamiltonians $\{{\hat H}\}$ that represent different
  physical systems, but which differ by a constant $C$ , and yet possess the same
  density  $\rho({\bf r})$, cannot be distinguished on the basis of the Hohenberg-Kohn theorem.

\pagebreak

\textbf{ B}.~  \textbf{Time-Dependent case}.\\

    We next extend the above conclusions to the time-dependent HK
    theorem. Consider again the Hooke's species, but in this case
    let us assume that the positions of the nuclei are time-dependent,
    i.e. $ {\bf R}_{j}={\bf R}_{j}(t)$. This could  represent, for
    example, the zero point motion of the nuclei. For simplicity we consider the spring constant
  strength to be the same ($k'$) for interaction with all the
  nuclei. The external
    potential energy $ v_{\mathcal{N}}({\bf r} t)$ for an arbitrary
 member of the species which  now is
  \begin{equation}
v_{\mathcal{N}}({\bf r} t)= \frac{1}{2} k'
\sum^{\mathcal{N}}_{j=1} ({\bf r}-{\bf R}_{j}(t))^{2},
     \end{equation}
   may then be rewritten as
   \begin{equation}
v_{\mathcal{N}}({\bf r} t)= \frac{1}{2} \mathcal{N}  k' r^{2}- k'
\sum^{\mathcal{N}}_{j=1}{\bf R}_{j}(t)\cdot {\bf r}+ \frac{1}{2}
k'
 \sum^{\mathcal{N}}_{j=1} {\bf R}^{2}_{j}(t),
    \end{equation}
    where at some initial time $t_{0}$, we have ${\bf
    R}_{j}(t_{0})={\bf R}_{j,0}$. (Note that a spatially uniform
    time-dependent field ${\bf F}(t) $ interacting only with the
    electrons could be further incorporated by adding a term ${\bf
    F}(t)\cdot{\bf r}$ to the external potential energy
    expression.) The Hamiltonian of an element of the species
    governed by the number  of nuclei $\mathcal{N}$ is then
   \begin{equation}
   \hat {H}_{\mathcal{N}}( {\bf r}_{1}  {\bf r}_{2} t)=\hat
   {H}_{\mathcal{N},0}-k' \sum^{\mathcal{N}}_{j=1}[{\bf R}_{j}(t)-{\bf R}_{j,0}]\cdot ({\bf
   r}_{1}+ {\bf r}_{2}) + C(k',\mathcal{N},t),
      \end{equation}
 where $\hat {H}_{\mathcal{N},0}$  is the time-independent Hooke's
 species Hamiltonian Eq.(21):
   \begin{equation}
\hat {H}_{\mathcal{N},  0}=\hat {H}_{\mathcal{N}}(k'),
\end{equation}
and the time-dependent function
 \begin{equation}
 C(k',\mathcal{N},t)= k' \sum^{\mathcal{N}}_{j=1}[{\bf
 R}^{2}_{j}(t)-{\bf R}                            _{j,0}^{2}].
\end{equation}
Note that the function $C(k',\mathcal{N},t)$ contains physical
information about the system: in this case, about the motion of
the nuclei  about their equilibrium positions.  It also
differentiates between
the different elements of the species.\\

  The solution of the time-dependent Schr{\"o}dinger  equation$(\hat {H}_{\mathcal{N}}(t)\Psi(t)= i \partial
  \Psi(t)/\partial t)$ employing the Harmonic Potential Theorem
  \cite{7} is
\begin{equation}
 \Psi({\bf r}_{1} {\bf r}_{2}t)=exp \{-i \phi(t) \} exp[-i \{E_{\mathcal{N},0} t- 2S(t)-2
 \frac{d {\bf z}}{dt}\cdot {\bf y} \}] \Psi_{0}( \overline{{\bf
 r}_{1}} \;\overline{{\bf r}_{2}}),
\end{equation}
where $\overline{{\bf r}_{i}}={\bf r}_{i}-{\bf z}(t),  {\bf
y}=({\bf r}_{1}+{\bf r}_{2})/2$,
\begin{equation}
S(t)=\int^{t}_{t_{0}}[\frac{1}{2}  \dot{{\bf z}}(t')^{2}-
\frac{1}{2}   k  {\bf z}(t')^{2}] dt',
\end{equation}
the shift ${\bf z}(t)$ satisfies the classical harmonic oscillator
equation
\begin{equation}
\ddot{{\bf z}}(t)+ k {\bf z}(t)-k' \sum^{\mathcal{N}}_{j=1}[{\bf
 R}_{j}(t)-{\bf R}_{j,0}]=0,
\end{equation}
where the additional phase factor $\phi(t)$ is due to the function
$C(k',\mathcal{N},t)$,
\begin{equation}
   \phi(t)=\int^{t}_{t_{0}} C(k',\mathcal{N},t')dt',
\end{equation}
and where at the initial time $\Psi({\bf r}_{1} {\bf r}_{2}
t_{0})=\Psi_{0}$ which satisfies $\hat {H}_{\mathcal{N},
 0}\Psi_{0} =E_{\mathcal{N},  0} \Psi_{0}$. Thus, the wave function
$\Psi({\bf r}_{1} {\bf r}_{2} t)$ is the time-independent solution
shifted by a time-dependent function ${\bf z}(t)$, and multiplied
by a phase factor. The explicit contribution of the function
$C(k',\mathcal{N},t)$ to this phase has been separated out. The
phase factor cancels out in the determination of the density
$\rho({\bf} t)=\langle \Psi(t)|\hat {\rho}|\Psi(t)\rangle=\rho
({\bf r}-{\bf z}(t))$ which is the initial time-independent
density $\rho ({\bf r} t_{0})= \rho _{0}({\bf r})$ displaced by
${\bf z}(t)$.\\

 As in the time-independent case, the `degenerate Hamiltonians' $\hat {H}_{\mathcal{N}}( {\bf r}_{1}  {\bf
 r}_{2} t)$ of the time-dependent Hooke's species can each be made
 to generate the \textit{\textbf{same}} density $\rho ({\bf r} t)$ by
 adjusting the spring constant $k'$ such that $\mathcal{N}k'=k$, and provided the density at the
 initial time $t_{0}$ is the \textit{\textbf{same}}. The latter is readily
 achieved as it constitues the time-independent Hooke's species
 case discussed previously.\\

 Thus, we have a set of Hamiltonians describing different physical
 systems but which can be made to generate the same density $\rho ({\bf r}
 t)$. These Hamiltonians differ by the function
 $C(k',\mathcal{N},t)$ that contains information which
 differentiates between them. In such a case, the density $\rho ({\bf r}
 t)$ cannot distinguish between the different Hamiltonians.\\

 \textit{\textbf{Corollary}}: Degenerate time-dependent
  Hamiltonians $\{ {\hat H}(t)\}$ that
 represent different physical systems, but which differ by a purely
 time-dependent function $C(t)$, and which all yield the same
 density $\rho ({\bf r}t)$, cannot be distinguished on the basis of the Runge-Gross theorem.\\

\section{III. ~ENDNOTE }
  The proof of the HK theorem is general in that it is valid for
  \textit{\textbf{arbitrary}} local form  ( Coulombic, Harmonic, Yukawa,
  oscillatory, etc.) of external potential energy $v({\bf r})/v({\bf r}
  t)$. (In the time-dependent case, there is the restriction that
  $v({\bf r} t)$ must be Taylor-expandable about some initial time
  $t_{0}$.) For their proof, HK/RG considered the case of potential
  energies , and hence Hamiltonians, that differ by an additive
  constant $C$/function $C(t)$ to be equivalent:
\begin{equation}
v({\bf r})/v({\bf r} t)- v'({\bf r})/v'({\bf r} t)=C/C(t).
\end{equation}
By equivalent is meant that the density $\rho ({\bf r} )/\rho ({\bf r}
t)$ is the same. The fact that the constant $C$/function $C(t)$ is
\textit{\textbf{additive}} means that although the Hamiltonians
differ, the physical system, however remains the
\textit{\textbf{same}}. The theorem then shows that there is a
one-to-one correspondence between a physical system (as described
by all these equivalent Hamiltonians), and  the corresponding density $\rho ({\bf
r} )/\rho ({\bf r} t)$. The relationship between the basic
Hamiltonian ${\hat H}/{\hat H}(t)$ describing a particular system
and the density $\rho ({\bf r} )/\rho ({\bf r} t)$ is bijective or
fully invertible. This case considered by HK/RG is shown
schematically in Fig. 1 in which the invertibility is indicated by
the double-headed arrow.\\

\begin{figure}
 \includegraphics[bb=60 202 791 534, angle=0.5, scale=0.7]{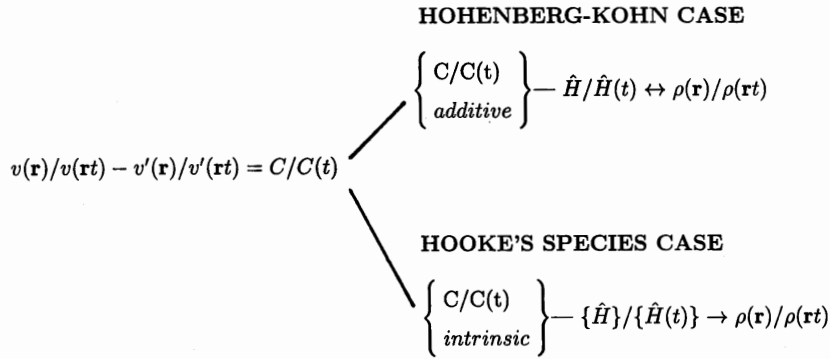}
 \caption{A schematic representation of the Hohenberg-Kohn theorem and its corollary.\label{}}
 \end{figure}

The case of a set of degenerate Hamiltonians $\{{\hat H}\}/\{{\hat
H}(t)\}$ that differ by a constant  $C$/function $C(t)$  that is
\textit{\textbf{intrinsic}} such that the Hamiltonians represent
\textit{\textbf{different}} physical systems while yet
\textbf{\textit{all}} possessing the same density $\rho ({\bf r}
)/\rho ({\bf r} t)$, was not considered by HK/RG. In such a case,
 the density
\textit{\textbf{cannot }} uniquely determine the Hamiltonian, and
therefore \textit{\textbf{cannot }}differentiate between the
different physical systems. This case, also shown schematically in Fig.1, corresponds to
 the Hooke's species.   The relationship between the set of
Hamiltonians $\{{\hat H}\}/\{{\hat H}(t)\}$ and the density $\rho
({\bf r} )/\rho ({\bf r} t)$ which is not invertible is indicated by the single-headed
arrow. \\

We conclude by noting that the Hooke's species, in both the time-independent and time-dependent
cases, does not contitute a counter example to the HK/RG theorem. The reason for this is that
the proof of the HK/RG theorem is \textit{\textbf{independent}} of whether the constant $C$/ function $C(t)$ is
additive or intrinsic. The Hamiltonians in either case still differ by a constant $C$/ function $C(t)$.
 A counter example would be one in which Hamiltonians that differ by more than a constant $C$/ function $C(t)$
have the same density $\rho ({\bf r} )/\rho ({\bf r}
t)$.









\subsection{}
\subsubsection{}

\begin{acknowledgments}
\textbf{ACKNOWLEDGMENT}\\

  We acknowledge invaluable discussions with Lou Massa and Ranbir
  Singh. This work was supported in part by the Research Foundation of the
  City University of New York.
\end{acknowledgments}
\pagebreak

\begin{references}
\bibitem{1} P. Hohenberg and W.Kohn, Phys. Rev. \textbf{136B}, 864(1964).
\bibitem{2}  E. Runge and E. K. U. Gross, Phys. Rev. Lett. \textbf{52}, 997(1984).
\bibitem{3}   T. Kato, Comm. Pure. Appl. Math \textbf{10}, 151(1957).
\bibitem{4} S. Kais, D. R. Herschbach, and R. D. Levine, J. Chem.
Phys. \textbf{91}, 7791(1989); M. Taut, Phys. Rev. A \textbf{48},
3561(1993).
\bibitem{5} Z. Qian and V. Sahni, Phys. Rev. A. \textbf{57}, 2527(1998).
\bibitem{6} S. Kais, D. R. Herschbach, N. C. Handy, C. W. Murray,
and G. J. Laming, J. Chem. Phys. \textbf{99}, 417(1993).
\bibitem{7} J. F. Dobson, Phys. Rev. Lett. \textbf{73}, 2244(1994).

\end{references}
\nopagebreak

\end{document}